\documentstyle[12pt]{article}\textheight 230mm\textwidth 150mm
\newcommand{\be}{\begin{eqnarray}}
\newcommand{\ee}{\end{eqnarray}}
\newcommand{\eel}[1]{\label{#1}\end{eqnarray}}
\newcommand{\beq}{\begin{quote}}
\newcommand{\eq}{\end{quote}}
\newcommand{\ben}{\begin{enumerate}}
\newcommand{\een}{\end{enumerate}}
\newcommand{\r}[1]{(\ref{e:#1})}
\newcommand{\lra}{{\leftrightarrow}}
\newcommand{\Lra}{{\Leftrightarrow}}
\newcommand{\dx}{\dot{x}}
\newcommand{\dga}{\dot{\gamma}}
\newcommand{\tp}{\tilde{p}}
\newcommand{\bp}{{\bf p}}
\newcommand{\bett}{{\bf 1}}
\newcommand{\bga}{\mbox{\boldmath $\ga$}}
\newcommand{\vb}{{\cal h}}
\newcommand{\hb}{{\cal i}}
\newcommand{\ra}{{\rightarrow}}
\newcommand{\nn}{\nonumber}
\newcommand{\eg}{{\em e.g.\ }}
\newcommand{\ie}{{\em i.e.\ }}
\newcommand{\al}{\alpha}
\newcommand{\ga}{{\gamma}}
\newcommand{\la}{{\lambda}}
\newcommand{\ka}{{\kappa}}
\newcommand{\del}{{\delta}}
\newcommand{\Om}{\Omega}
\newcommand{\pet}{{\cal P}}
\newcommand{\dagg}{^{\dag}}

\newcommand{\bac}{\bar{c}}

\newcommand{\bak}{\bar{k}}

\newcommand{\bata}{\bar{\eta}}
\newcommand{\barho}{\bar{\rho}}
\newcommand{\bapet}{\bar{\pet}}

\newcommand{\halv}{\frac{1}{2}}
\newcommand{\kvart}{\frac{1}{4}}
\begin{document}
\begin{titlepage}
\noindent
G\"{o}teborg ITP 93-18\\
August 1993\\
\vspace*{17 mm}
\begin{center}{\LARGE\bf Proper BRST quantization\\of relativistic
particles.}\end{center} \begin{center}\vspace*{14 mm}
\begin{center}Robert Marnelius \\
\vspace*{8 mm}
{\sl Institute of Theoretical Physics\\
Chalmers University of Technology\\
S-412 96  G\"{o}teborg, Sweden}\end{center}
\vspace*{20 mm}
\begin{abstract}
Recently derived general formal solutions of a BRST quantization on inner
product spaces of
irreducible Lie group gauge theories are applied to trivial models and
relativistic
particle models for particles with spin $0, 1/2$ and $1$. In the process
general quantization rules
are discovered which make the formal solutions exact. The treatment also give
evidence that the formal
solutions are directly generalizable to theories with graded gauge symmetries.
For relativistic
particles  reasonable results are obtained although there exists no completely
Lorentz covariant
quantization of the coordinate and momenta on inner product spaces. There are
two inequivalent
procedures depending on whether or not the time coordinate is quantized with
positive or indefinite
metric states. The latter is connected to propagators.
\end{abstract}\end{center} \end{titlepage}

\newpage

\setcounter{page}{1}
\section{Introduction.}
 For a long time there has been a notorious difficulty to perform a unique and
well
defined BRST quantization of quantum mechanical systems like those describing
relativistic
particles, relativistic strings or any other field theory with nontrivial zero
modes.
These difficulties have their origin in the use of badly defined original state
spaces. For
instance, the solutions of a BRST quantization of particles and strings are
usually presented in
terms of momentum eigenstates which do not belong to an inner product space.
Such a result is only
possible if the projection to the BRST invariant subspace is obtained from a
state space which
also includes these momentum states. To achieve this one has  to work with
bilinear forms
constructed from the chosen state space and its dual, which in the above case
does not contain
the momentum eigenstates. Such a construction has been used in a series of
papers in which we have
used state spaces which are analytic in the momenta while the dual state space
is analytic in the
coordinates \cite{HMa,Hwang,RU,MO}. In this construction it was
noted that BRST invariance of the dual state space selects states which make
finite bilinear forms
with the BRST invariant momentum states. New light was cast on this procedure
in \cite{Ax,Aux} where
the allowed properties of general gauge fixing within a BRST scheme were
determined. It was shown
that the BRST invariant states may be further constrained by operators which
form BRST doublets
satisfying a closed algebra. In \cite{Aux} the relativistic particle and string
were explicitly
treated and it was shown exactly which constraints the dual BRST invariant
states have to satisfy
when one makes use of BRST invariant momentum states. These constraints are
noncovariant and point to
a difficulty with the above procedure involving the state space and its dual.

What one normally should require of the outcome of a BRST quantization
is that the BRST cohomology should yield a positive definite state space (a
Hilbert space).
However, this causes a severe difficulty within the above construction. In
order to satisfy this
condition one must be able to define an inner product for the genuine physical
states which belong to
the BRST cohomology. In \cite{MO}     it was suggested that a rather natural
inner product may be
constructed provided one makes use of dynamical Lagrange multipliers and
antighosts. However, this
proposal
 contained  some confusing technical difficulties and was therefore not
entirely correct.
A suggestion how these difficulties could be overcome was given in \cite{Gen}.
There it was
suggested that a correct treatment requires the use of spectral resolutions
appropriate for an
original state space being an inner product space.  In \cite{Simple,Gauge} it
was then
shown that a BRST quantization of general gauge models may always be performed
and formally solved in
a simple manner on inner product spaces and that the solutions modify the
proposal in \cite{MO}.

Presently it seems therefore as if the only way to get rid of all the
previous difficulties and ambiguities is to require the original state space in
a BRST quantization
to  be an inner product space, since  in this case  it looks as if we always
can
obtain a well defined  BRST quantization. The purpose of the present paper is
 to further explore this possibility for models which usually are difficult to
handle by a detailed
treatment of formal solutions of the type given in \cite{Simple,Gauge}. On our
way we will then find general rules how the original state space should be
chosen in order to have  a
well defined BRST quantization.

One point made in \cite{Gen} was that Hermitian coordinate and momentum
operators may be considered
to act within an inner product space, and that we may make use of a complete
set of coordinate and
momentum eigenstates in  a well defined manner although these states do not
belong to the inner
product space. How this works in the ordinary Hilbert space is well known.
However, in the case when
we have an indefinite metric state space it is not so well known that one may
still make use of a
complete set of eigenstates {\em provided} the eigenvalues are properly chosen.
The  hermitian
coordinate and momentum operators must have imaginary eigenvalues when the
corresponding state space
is indefinite \cite{Pauli}. Real eigenvalues are only consistent with  positive
definite state spaces
which is apparent from (assuming a positive measure $dx$) \be
&&\vb\phi|\phi\hb=\int dx\vb\phi|x\hb\vb x|\phi\hb=\int dx|\phi(x)|^2\geq0
\eel{e:1}
(In \cite{Gen} the possibility of using real eigenvalues and infinite normed
state spaces regularized
to antihermitian inner products was also considered.)

The implications of using an inner product space representation in a general
BRST quantization of an
arbitrary consistent model were summerized in section 5 of ref.\cite{RMa}
(extracted from \cite{HMa}).
There it was stated that in order for the genuine physical states to span a
Hilbert space:
\ben
\item{The number of constraints must be even in the chosen gauge theory.}
\item{The BRST charge $Q$ must be possible to write as
\be
&&Q=\del^{\dag}+\del
\eel{e:2} where
\be
&&\del^2=\del^{\dag 2}=[\del, \del^{\dag}]_+=0
\eel{e:3}}
\item{The genuine physical states may be chosen to have ghost number zero.}
\item{The genuine physical states may always be chosen to be determined by a
generalized Gupta-Bleuler
quantization.} \een
(Points $1$ and $3$ are well known properties.)

 These properties suggest that one in general should make use of dynamical
Lagrange multipliers as well as antighosts since this will always imply that we
have an even number
of constraints and that the ghost number zero is contained in the original
state space. Furthermore,
they will in general  allow for antiBRST invariance \cite{Hwa} which in turn
allows for larger OSP
invariances \cite{Ax}. The presence of the antighosts makes sure that we have
an even number of both
bosonic and fermionic ghosts, the first of which allows for the possibility to
have a Fock space
representation and the second of which avoids the problem with an odd number of
fermionic ghosts
\cite{Mar}. Indeed, in \cite{Simple,Gauge} it was shown that condition 2 may be
satisfied in
several different ways, at least for any Lie group theory, and that the general
solutions of the
BRST condition yields physical states which satisfy conditions 3 and 4  when
dynamical Lagrange
multipliers are introduced. (Decompositions like \r{2} with \r{3} have been
used before. See \eg
\cite{Slav}.)

The paper is organized as follows: In  section 2
the general properties found in \cite{Simple,Gauge} is briefly reviewed  with
some additional
 remarks and in section 3 and 4 the main quantization rules are obtained from a
 detailed treatment
of the abelian case. A detailed treatment of the spinless relativistic particle
is then given in
section 5 where the properties of state spaces of coordinate and momentum
states are further
clarified. The world-line supersymmetric spin-$\halv$ particle is  treated in
section 6, and the
$O(2)$-extended  world-line supersymmetric spin-$1$ particle is  treated in
section 7. In section 8
we give some final remarks. In an appendix the properties of the spinor states
for the
spin-$\halv$ particle are given.

\section{The general  case.}
\setcounter{equation}{0}
In \cite {Simple} it was proved that the  BRST charge of a general bosonic
gauge model with finite
number of degrees of freedom may be written in the form \r{2} with the
properties \r{3} provided
dynamical Lagrange multipliers and antighosts are introduced. The starting
point was the BRST charge
in the BFV form  \cite{BV} ($a, b, c =1,\ldots,m<\infty$)
\be
&&Q=\psi_a\eta^a-\frac{1}{2}iU_{bc}^{\;\;a}\pet_a
\eta^b\eta^c-\frac{1}{2}iU_{ab}^{\;\;b}\eta^a + \bapet_a\pi^a
\eel{e:201}
where the irreducible set of gauge generators $\psi_a$  satisfy
\be
&[\psi_a, \psi_b]_{-}=iU_{ab}^{\;\;c}\psi_c
\eel{e:202}
and where $U_{ab}^{\;\;c}$ are constants. The anticommuting ghosts $\eta^a$,
antighosts $\bata_a$ and
the commuting Lagrange multipliers $v^a$ satisfy the algebra (the nonzero part)
\be
&[\eta^a, \pet_b]_{+}=[\bata^a, \bapet_b]_{+}=\del^a_{\;b},\;\;\;[v^a,
\pi_b]_-=i\del^a_{\;b}
\eel{e:203}
All these variables are assumed to be hermitian, which is no restriction.
In \cite{Simple,Gauge} it was  shown that the charge \r{201}  may be
written in the form \r{2} where
\be
&&\del=c^{\dag a}\phi_a=\phi'_ac^{\dag a}
\eel{e:204}
where in turn the non-hermitian operators $\phi_a$ or $\phi'_a$
($=\psi_a+\ldots$)
satisfy the same Lie algebra as $\psi_a$, \ie \r{202}. $c^a$ is an expression
in terms of
ghosts, antighosts and Lagrange multipliers in \cite{Simple}. In \cite{Gauge}
the construction
involves gauge fixing variables instead of Lagrange multipliers. By means of a
bigrading which implies
\cite{RMa}
\be &&Q|ph\hb=0\; \Lra \;\del|ph\hb=0,\;\;\del^{\dag}|ph\hb=0
\eel{e:205}
the BRST condition was then shown to be naturally solved by
\be
&&c^a|ph\hb=0,\;\;\;\phi_a|ph\hb=0
\eel{e:206}
or
\be
&&c^{\dag a}|ph\hb=0,\;\;\;{\phi'_a}^{\dag}|ph\hb=0
\eel{e:207}
(or possibly a mixture of these conditions). Other solutions of \r{205} are in
general zero norm
states \cite{Gauge}.

 The formal solutions of \r{206} and \r{207} were shown to be
\be
&&|ph\hb_{\al}=e^{\al[\rho, Q]}|\Phi\hb
\eel{e:208}
where $|\Phi\hb$ is a BRST invariant state and where $\al$ is a strictly
positive or a strictly
negative real constant for \r{206} and \r{207} respectively. In \cite{Simple}
we have \be
&&\rho\equiv\pet_av^a.
\eel{e:209}
and the state $|\Phi\hb$  satisfies the conditions
\be
&&\eta^a|\Phi\hb=\pi_a|\Phi\hb=0
\eel{e:210}
which makes it BRST invariant. These conditions are trivially  solved. In
\cite{Gauge} we have the
formal solution \r{208} with
 \be
&&\rho\equiv\bata_a\chi^a.
\eel{e:2101}
where $\chi^a$ are gauge choices or gauge fixing variables for $\psi_a$
($[\chi^a, \psi_b]_-$ must
have an inverse). The state $|\Phi\hb$  satisfies here
\be
&&\pet_a|\Phi\hb=\bapet_a|\Phi\hb=(\psi_a + iU_{ab}^{\;\;b})|\Phi\hb=0
\eel{e:2102}
which  makes it  BRST invariant. The last conditions are just those of a Dirac
quantization which are
not trivial to  solve in general. The norm of $|ph\hb_{\al}$ should be
independent of the value of
$\al$ but can depend on the sign of $\al$ since the two signs requires
different representations of
the original state space $\Om$.

In order for \r{208} to be true solutions we must show
that they belong to an inner product space. In other words
although $|\Phi\hb$ does not belong to an inner product
 space $|ph\hb_{\al}$ must do. As a consequence
\be
&&|ph\hb_{\al}=|\Phi\hb+Q|\chi\hb
\eel{e:211}
does not imply that we may through away $Q|\chi\hb$!

The philosophy of the above approach is to start with a large inner product
space $\Om$ and project
out a genuine physical state space $\Om_{ph}\subset\Om$ which then also is an
inner product space. In
a {\em canonical theory} (defined in \cite{HMa}) the BRST condition \r{205}
yields
a unique set of
solutions for a given original state space. If the BRST condition leads to
several different
conditions of the form \r{206} etc. with non-trivial solutions then there exist
several different
physical state spaces $\Om_{ph},\Om'_{ph},\cdots\subset\Om$. If $\Om$ really is
an inner product
space then $\Om_{ph},\Om'_{ph},\cdots$ may be combined into one physical state
space which, however,
means that the BRST cohomology contains ghost excitations which in turn implies
that it cannot be a
positive  state space. Hence,   a satisfactory theory must be a canonical one
(see also section $7$).
Now the solutions \r{208} are obtained in a formal manner without using that
$\Om$ is an inner product
space. Thus, in our procedure as described above we do not specify in advance
how the ghosts and
Lagrange multipliers are to be represented in $\Om$. Instead we let the
condition that $\Om_{ph}$
must belong to an inner product space specify the appropriate representation in
$\Om$. Through the
 examples to be considered this condition leads to the proposal of the
following general rules:
\ben
\item Lagrange multipliers must be quantized with opposite metric states to the
unphysical variables
in the matter space (the variables which $\psi_a$ eliminate).
\item Bosonic ghosts and antighosts must be quantized with opposite metric
states.
\een
Only if these conditions are satisfied can $|ph\hb$ belong to an inner product
space. However, in
general we must also restrict the range of the Lagrange multipliers.
It is the choice of representation of $\Om$ which determines whether we have
the solutions
$|ph\hb_+$  or $|ph\hb_-$. We cannot have both since $_+\vb
ph|ph\hb_-=\vb\Phi|\Phi\hb$ is undefined.
However, it is natural to require that $|ph\hb_+$  and $|ph\hb_-$ yield
equivalent physics when both
possibilities are allowed. We require therefore their
norms to be exactly the same. In the following examples this condition is shown
to partly specify
the representation of $\Om$ and sometimes to restrict  the ranges of the
Lagrange multipliers.
The method can be used to determine a canonical theory in which case $\Om$ is
an inner product
space for which the BRST condition yields a unique physical state space. In
general this condition
severely restricts the possible forms of $\Om$ and the model.

\section{The general abelian case.}
\setcounter{equation}{0}
Consider the BFV charge \r{201} in the abelian case when $U_{ab}^{\;\;c}=0$,
\ie
\be
&&Q=\psi_a\eta^a + \bapet_a\pi^a
\eel{e:301}
where $\psi_a$ satisfies $[\psi_a, \psi_b]_{-}=0$. From the general formulas of
\cite{Simple} this
charge is   trivially  written as (see also \cite{Slav})
 \be
&&Q=c^{\dag a}\phi_a+\phi_a^{\dag}c^{ a}
\eel{e:302}
where
\be
&&\phi_a=\psi_a-i\pi_a,\;\;\;c^a=\halv(\eta^a-i\bapet^a)
\eel{e:303}
Hence, $Q$ is of the form \r{2} with
\be
&&\del=c^{\dag a}\phi_a
\eel{e:304}
Thus, if one imposes the bigrading leading to \r{205}, the BRST condition
$Q|ph\hb=0$ is either
solved by
\be
&&c^a|ph\hb=0,\;\;\;\phi_a|ph\hb=0
\eel{e:305}
or
\be
&&c^{\dag a}|ph\hb=0,\;\;\;\phi_a^{\dag}|ph\hb=0
\eel{e:306}
(or a mixture of these cases).
The solutions may be written
\be
&&|ph\hb_{\pm}=e^{\pm[\rho,
Q]}|\Phi\hb=e^{\pm(\psi_av^a+i\pet_a\bapet^a)}|\Phi\hb
\eel{e:307}
where $\rho\equiv \pet_av^a$ and where $|\Phi\hb$ satisfies \r{210}.
($|ph\hb_{+}$ solves \r{305} and $|ph\hb_{-}$ solves \r{306}.)

Now the decomposition \r{302} is not unique. Instead of \r{303} we may  also
choose
\be
&&\phi_a=\psi_a-i\frac{\pi_a}{\al},\;\;\;c^a=\halv(\eta^a-i\al\bapet^a)
\eel{e:308}
for any real parameter $\al\neq0$. The corresponding solutions are then
\be
&&|ph\hb_{\pm\al}=e^{\pm\al[\rho, Q]}|\Phi\hb
\eel{e:309}
where $|\Phi\hb$ still satisfies \r{210}. In fact, \r{307} and \r{309} are
connected by a unitary
transformation \cite{Simple}
\be
&&(\pi_a, v^a) \;\rightarrow\; (\frac{\pi_a}{\al}, \al v^a),\;\;\;(\bapet^a,
\bata_a) \;\rightarrow\;
(\al\bapet^a, \frac{\bata_a}{\al})
\eel{e:310}
We may therefore consider \r{309} for arbitrary real $\al\neq 0$. Consider,
therefore, in
particular $|ph\hb_{\pm\halv}$. Its inner product is given by
\be
&&_{\pm\halv}\vb ph|ph\hb_{\pm\halv}=\frac{1}{m!}(\pm
i)^m\vb\Phi|(\pet_a\bapet^a)^me^{\pm\psi_av^a}|\Phi\hb
 \eel{e:311}
which only is nonzero for those $|\Phi\hb$-states which also satisfy
\be
&&\bata_a|\Phi\hb=0
\eel{e:312}
in addition to \r{210}. Eqns \r{210} and \r{312} reduce $|\Phi\hb$ to
\be
&&|\Phi\hb=|\phi\hb|0\hb_{\pi}|0\hb_{\eta\bata}
\eel{e:313}
where $\pi_a|0\hb_{\pi}=0$,
$\eta^a|0\hb_{\eta\bata}=\bata_a|0\hb_{\eta\bata}=0$ and where
$|\phi\hb$ only depends on the matter variables, which  are involved in \eg
$\psi_a$, but not on
ghosts and Lagrange multipliers. Using the convention of \cite{Mar} we get
\be
&&\frac{1}{m!}(\pm
i)^m\,_{\eta\bata}\vb0|(\pet_a\bapet^a)^m|0\hb_{\eta\bata}=(\pm1)^m C,\;\;C=\pm
1
\eel{e:314}
Hence, \r{311} reduces to
\be
&&_{\pm\halv}\vb ph|ph\hb_{\pm\halv}=(\pm1)^m C
\vb\phi|\,_{\pi}\vb0|e^{\pm\psi_av^a}|0\hb_{\pi}|\phi\hb
 \eel{e:315}
 The middle
term may be calculated by means of the spectral representation of the Lagrange
multipliers $v^a$.
As was already mentioned in \cite{Gen} the only way to obtain a finite norm is
to quantize $v^a$
 with opposite metric states to the variable which
$\psi_a$ eliminates. In particular, when $|\phi\hb$ belongs to a positive
definite state space (a
Hilbert space)  all Lagrange multipliers $v^a$ must be quantized with
indefinite metric states for
which  the appropriate eigenvalues are imaginary  \cite{Pauli,Gross,HeT,Gen}.
We have therefore the
spectral representation
\be
&&v^a|iu\hb=iu^a|iu\hb
\eel{e:316}
which  satisfies
\be
&&\vb -iu|iu'\hb=\del^m(u'+u)\nonumber\\
&&\vb-iu|\equiv(|iu\hb)^{\dag}
\eel{e:3161}
Hence, we have
\be
&&\vb iu|iu'\hb=\del^m(u'-u)\nonumber\\
&&\int\!d^mu|iu\hb\vb iu|=\bett=\int\!d^mu|-iu\hb\vb -iu|
\eel{e:317}
(Notice that the range of $u$ must be symmetric: $-a\leq u\leq a$.)
When this is inserted into \r{315} we obtain
\be
&&_{\pm\halv}\vb ph|ph\hb_{\pm\halv}=(\pm1)^m C
\vb\phi|\int\!d^mue^{\pm i\psi_au^a}|\phi\hb=\nn\\&&=(\pm1)^m
C(2\pi)^m\int\!dx\del^m(\psi_a)|\phi(x)|^2
 \eel{e:318}
where we also have introduced a complete set of states $|x\hb$ in the matter
space,
\ie $\int\! dx|x\hb\vb x|=\bett$. The last line in \r{318} is formal.

The sign difference between the norms of $|ph\hb_+$ and $|ph\hb_-$  when we
have an odd number of
constraints ($m$ odd) may also be directly understood from \r{305} and \r{306}.
Obviously the
nontrivial solutions of \r{305} and \r{306} satisfy
\be
&&c^a|ph\hb_+=k_a|ph\hb_+=0\nn\\
&&c^{\dag a}|ph\hb_-=k\dagg_a|ph\hb_-=0
\eel{e:3181}
respectively where
\be
&&[c^a, k\dagg_b]_+=\del^a_b
\eel{e:3182}
This means that \r{305} picks up the ghost vacuum
\be
&&c^a|0\hb=k_a|0\hb=0
\eel{e:3183}
while \r{306} picks up
\be
&&c^{\dag a}\bar{|0\hb}=k\dagg_a\bar{|0\hb}=0
\eel{e:3184}
Both vacua have ghost number zero. In \cite{Mar} (eqn (4.18)) it is shown that
\be
&&\bar{\vb0|}\bar{0\hb}=(-1)^m\vb0|0\hb
\eel{e:3185}
when $\vb0|0\hb=\pm1$.

The condition that $|ph\hb_+$ and $|ph\hb_-$ should yield equivalent physics
requires from \r{315}
and \r{318} that we either choose different ghost representations or different
norms of the bosonic
vacuum state in the two cases when $m$ is odd. (In the last line of \r{318} we
have assumed that the
bosonic vacuum has positive norm.) The last possibility is probably the most
natural choice since
$|ph\hb_+$ and $|ph\hb_-$
involve different vacuum states. It is probably natural that even these bosonic
vacua should be
related according to \r{3185}.

\section{A trivial model.}
\setcounter{equation}{0}
In order to make the above description even more explicit we consider the
simple abelian case when
$\psi_a=p_a$ where $p_a$ are canonical momenta to some coordinates $x^a$
satisfying
\be
&&[p_a, x^b]_-=i\del^b_{\;a}
\eel{e:319}
In this case we  have either a spectral representation in which $x^a, p_a$ have
real eigenvalues and
$v^a, \pi_a$ imaginary ones, or vice versa, or a mixture. In all cases \r{315}
reduces to
\be
&&_{\pm\halv}\vb ph|ph\hb_{\pm\halv}=(\pm1)^m C
\int\!d^mp\, d^mv e^{\pm ip_av^a}|\phi(p)|^2=\nn\\&&=(\pm1)^m C
(2\pi)^m|\phi(0)|^2
 \eel{e:320}
after insertion of the completeness relations for the eigenstates of $p_a$ and
$v^a$. Thus, the
physical state space, which here only is  a vacuum state, is either positive
definite or negative
definite and this sign depends on the choice of ghost representation of the
original state space
$\Om$ and whether we have the solutions $|ph\hb_+$ or $|ph\hb_-$. (It does also
depends on the
vacuum normalization of the bosonic variables \r{319} which is assumed to be
positive in \r{320}.)

The meaning of the statement that $x^a$ and $v^a$ are quantized with positive
or an indefinite
metric refers to a choice of oscillator basis. For instance, if the indices are
assumed to be raised
or lowered with an Euclidean metric the oscillators
\be
&&a^a=\frac{1}{\sqrt{2}}(x^a+ip^a),\;\;\;b^a=\frac{1}{\sqrt{2}}(v^a+i\pi^a)
\eel{e:321}
span a positive definite state space for $x^a$ and $p_a$, and an indefinite one
for $v^a$ and
$\pi_a$ while the oscillator basis spanned by
\be
&&a^a=\frac{1}{\sqrt{2}}(x^a-ip^a),\;\;\;b^a=\frac{1}{\sqrt{2}}(v^a-i\pi^a)
\eel{e:322}
has the opposite properties. (This follows from \r{203} and \r{319}.) Thus, the
results \r{320}
requires an original state space spanned by a Fock basis with equally many
positive and indefinite
oscillators.

The appropriate Fock space is here the state space spanned by the oscillators
\be
&&\phi_a=p_a-i\pi_a,\;\;\;\xi_a=\halv(ix_a-v_a),\;\;\;[\xi_a,
\phi_b^{\dag}]_-=\del_{ab}
\eel{e:3221}
since its vacuum state
\be
&&\phi_a|0\hb=0,\;\;\;\xi_a|0\hb=0
\eel{e:3222}
will be picked up by the condition \r{305}. The
 corresponding diagonal basis of \r{3221} is spanned by \eg
\be
&&a_a\equiv\halv(\xi_a+\phi_a),\;\;\;b_a\equiv\halv(\xi_a-\phi_a)
\eel{e:3223}
where $a_a$ and $b_a$ are positive and indefinite oscillators respectively.
Thus, the
oscillator basis \r{3221} is equivalent to a basis with equally many positive
as indefinite
oscillators which was exactly the requirement for \r{320}.
Alternatively we may span the original state space by the oscillators \r{3221}
with $\phi_a$ and
$\xi_a$ counted as creation operators. The vacuum state is then
\be
&&\phi\dagg_a\bar{|0\hb}=0,\;\;\;\xi\dagg_a\bar{|0\hb}=0
\eel{e:3224}
which will be picked up by eqn \r{306}. Thus, it is the choice of Fock space
representation of the
original state space which determines whether $|ph\hb_+$ or $|ph\hb_-$ are
non-trivial solutions.
Notice also that the wave function representation of $|0\hb$ and $\bar{|0\hb}$,
$\phi_0(p,v)\equiv\vb
p,v|0\hb$ and $\bar{\phi_0}(p,v)\equiv\vb p,v\bar{|0\hb}$, requires imaginary
eigenvalues of either
$p_a$ or $v^a$ if $|0\hb$ and $\bar{|0\hb}$ are to be normalizable
($\phi_0(p,v)\propto e^{p_av^a}$).
The condition that $|ph\hb_+$ and $|ph\hb_-$ should have the same norms is
naturally satisfied if
the normalization of the vacua \r{3222} and \r{3224} are related by
\be
&&\vb 0|0\hb=(-1)^m\bar{\vb0|}\bar{0\hb}
\eel{e:3225}
(cf. \r{3185}).

 In order to get a first hint of
what a graded gauge group requires within the framework of \cite{Simple}, we
interchange the meaning
of the variables in $Q=p_a\eta^a+\pi_a\bapet^a$. Let therefore $\eta^a$ be
fermionic gauge generators
and $p_a$ bosonic ghosts, $\bata^a$ Lagrange multipliers and $v^a$ antighosts.
The BRST invariant
solutions corresponding to \r{309} are then \be
&&|ph\hb_{\pm\al}=e^{\pm\al[\barho, Q]}|\Phi'\hb
\eel{e:323}
where ($\barho=x^a\bata_a$)
\be
&&[\barho, Q]_+=x^a\pi_a-i\bata_a\eta^a
\eel{e:324}
and where $|\Phi'\hb$ satisfies
\be
&&\bapet^a|\Phi'\hb=p_a|\Phi'\hb=0
\eel{e:325}
(Actually, \r{323} satisfies the same conditions as \r{309} with $\al\ra
1/{\al}$.)
The inner product of \r{323} is (cf.\r{315})
\be
&&_{\pm\halv}\vb ph|ph\hb_{\pm\halv}=(\pm1)^m C
\vb\phi'|\,_{p}\vb0|e^{\pm\pi_ax^a}|0\hb_{p}|\phi'\hb
 \eel{e:326}
which may be calculated by means of the spectral representations of $x^a$ and
$\pi_a$. The only way
to get a constant as in \r{320} is then to impose the condition that {\em
bosonic ghosts and
antighosts must be quantized with opposite metric states.} With this rule
\r{326} reduces to
\be
&&_{\pm\halv}\vb ph|ph\hb_{\pm\halv}=(\pm1)^m C
\int\!d^m\pi\, d^mx e^{\pm i\pi_ax^a}|\phi'(\pi)|^2=\nn\\&&=(\pm1)^m C
(2\pi)^m|\phi'(0)|^2
 \eel{e:327}
which is exactly equivalent to \r{320} which we should have since we have used
the same BRST charge
and the same state representation. Notice that in this case \r{3221} spans the
conventional
non-diagonal ghost state  representation and the vaccum \r{3222} is a ghost
vaccum with zero ghost
number.

{}From the point of view of the gauge fixing procedure of \cite{Gauge} the
physical states \r{323} are
solutions of the BRST condition with $p_a$ still counted as a constraint
variable and $\eta^a$ as
a ghost. $x^a$ in $\bar{\rho}$ is then a gauge fixing variable to $p_a$.
Similarly \r{307} is a
solution with $\eta^a$ as a constraint variable and $p_a$ as a ghost. $\pet_a$
in $\rho$ is then gauge
fixing variable to $\eta^a$.

\section{The spinless relativistic particle.}
\setcounter{equation}{0}
A  manifestly Lorentz covariant quantization of relativistic particles and
strings requires an
original state space for which there is a basis and a spectral representation
which transform
manifestly covariantly under Lorentz transformations. The problem is that these
properties cannot
be satisfied by an original state space which also is an inner product space
\cite{Gen}. In order to
see this consider the manifestly Lorentz covariant and hermitian coordinate and
momentum operators
$X^{\mu}$ and $P^{\mu}$ satisfying \be
&&[X^{\mu}, P^{\nu}]_-=i\eta^{\mu\nu}
\eel{e:401}
where $\eta^{\mu\nu}$ is a space-like Minkowski metric, \ie diag
$\eta^{\mu\nu}=(-1,+1,+1,+1)$. The
corresponding manifestly covariant oscillator representation \be
&&a^{\mu}=\frac{1}{\sqrt{2}}(X^{\mu}+iP^{\mu})
\eel{e:402}
satisfies
\be
&&[a^{\mu}, a^{\nu \dag}]_-=\eta^{\mu\nu}
\eel{e:403}
which means that the space components yields positive metric states while the
time components yields
indefinite metric states. However, an inner product space built on this basis
is not consistent with a
spectral representation with real eigenvalues, \ie states $|x\hb$ and $|p\hb$
satisfying
\be
&&X^{\mu}|x\hb=x^{\mu}|x\hb,\;\;\;P^{\mu}|p\hb=p^{\mu}|p\hb
\eel{e:404}
where $x^{\mu}$ and $p^{\mu}$ are real and with the completeness
relations
\be
&&\int\!d^4x|x\hb\vb x|=\bett,\;\;\; \int\!d^4p|p\hb\vb p|=\bett
\eel{e:405}
Instead it turns out that a Hilbert space topology requires \r{404} with
imaginary
eigenvalues for $X^0$ and $P^0$ in conjunction with the oscillator basis
\r{402}-\r{403}, which means
that the measures in \r{405} are not manifestly Lorentz invariant \cite{Gen}.
Thus, there are two
possibilities to represent $X^{\mu}$ and $P^{\mu}$ on an inner product space:
Either we choose the
covariant oscillator basis \r{402}-\r{403} with indefinite metric states and
imaginary eigenvalues
for $X^0$ and $P^0$, or we choose a non-covariant basis with positive metric
states (interchange
$a^0\;\lra\;a^{0\dag}$) but with real
 eigenvalues for $X^0$ and $P^0$ and
Lorentz invariant measures in \r{405}. (In \cite{Gen} an effort was made to
combine a Lorentz
covariant basis and real eigenvalues in \r{404} in an inner product space.
However, the resulting
inner products turned  out to be non-hermitian.) In the following we shall
perform a BRST
quantization of the relativistic particle for the above two choices of an
original inner product
space.

A spinless relativistic particle with mass $m$ is \eg given by  the Lagrangian
and Hamiltonian
\be
&&L=\frac{1}{2v}{\dot{x}}^2-\halv mv,\;\;\;H=\halv(p^2+m^2)v
\eel{e:406}
where $v$ is the Lagrange multiplier (usually called the einbein variable). The
corresponding
BRST invariant formulation gives rise to the BRST charge operator
\be
&&Q=\halv(P^2+m^2)\eta+\pi\bapet
\eel{e:407}
which may be written in the form \r{302} with \eg
\be
&&\phi_a=\phi=\halv(P^2+m^2)-i\pi,\;\;\;c^a=c=\halv(\eta-i\bapet)
\eel{e:408}
The solutions of \r{305} and \r{306} are from \r{307}
\be
&&|ph\hb_{\pm}=e^{\pm\halv(P^2+m^2)v\pm i\pet\bapet)}|\phi\hb
\eel{e:409}
where $|\phi\hb$ satisfies
\be
&&\pi|\phi\hb=\eta|\phi\hb=0
\eel{e:410}
The inner product of \r{409} is
\be
&&_{\pm}\vb ph|ph\hb_{\pm}=\pm
2i\vb\phi| e^{\pm(P^2+m^2)v}\pet\bapet|\phi\hb
 \eel{e:411}
where $|\phi\hb$ now may be chosen to be  in the form \r{313}. We get then
\be
&&_{\pm}\vb ph|ph\hb_{\pm}=\pm
2\vb\phi|e^{\pm(P^2+m^2)v}|\phi\hb
 \eel{e:412}

We have now to choose between the two possibilities mentioned above. Consider
first the case when
$X^0$ and $P^0$ are represented by a positive definite state space. In this
case the Lagrange
multiplier $v$ must be quantized with indefinite metric states which means
that we have to use a
spectral representation with imaginary eigenvalues of $v$ which we choose to be
$iu$. Inserting
the appropriate completeness relations into \r{412} we get (choosing the plus
sign by a choice of
vacuum normalization) \be
&&_{\pm}\vb ph|ph\hb_{\pm}=2\int\!d^4p\, du\, e^{\pm
i(p^2+m^2)u}|\phi(p)|^2=\nn\\
&&=4\pi\int\!d^4p\, \del(p^2+m^2)|\phi(p)|^2
 \eel{e:413}
where we have chosen the vacuum normalizations so that the norms are positive.
This is a correct
inner product for a free spinless relativistic particle apart from the fact
that it contains both
positive and negative energy solutions. (Within the gauge fixing procedure of
\cite{Gauge}, the
positive and negative energy solutions enter with opposite norms.)

If we instead make use of a manifestly covariant oscillator basis, then $X^0$
and $P^0$ are
represented by indefinite metric states and their spectral representation
requires imaginary
eigenvalues. In this case \r{412} becomes (again choosing the plus sign by a
choice  of vacuum
normalization) \be
&&_{\pm}\vb ph|ph\hb_{\pm}=2\int\!d^4p\,dv\,e^{\pm
(p^2+m^2)v}\phi^*(-p^0,\bp)\phi(p^0,\bp)
 \eel{e:414}
where $p^0$ now is an Euclidean energy, \ie $p^2+m^2$ is positive definite and
$d^4p$ is a Euclidean
measure. In order for this expression to be finite, the range of $v$ has to be
restricted: In
$|ph\hb_{+}$ $v$ has to have a finite maximal value, and in $|ph\hb_{-}$ a
finite minimal value. The
condition that $|ph\hb_{\pm}$ must yield equivalent results requires then
 \be
&&\;\;\;0\leq v<\infty\; \mbox{in}\;|ph\hb_{+}\nn\\
&&-\infty<v\leq0 \;\mbox{in}\; |ph\hb_{-}
\eel{e:415}
since only for these choices do we get the same inner products for $|ph\hb_{+}$
and $|ph\hb_{-}$ in
\r{414}, namely
 \be
&& _{\pm}\vb
ph|ph\hb_{\pm}=2\int\!d^4p\,\frac{\phi^*(-p^0,\bp)\phi(p^0,\bp)}{p^2+m^2}
\eel{e:416}
which is positive definite only if $\phi$ is an even function of $p^0$ (or an
odd function if we
choose a minus sign in \r{412}) In such a case this looks like a norm for a
free Euclidean field.
A Lorentz invariant way to make $\phi$ an even function of $p^0$ is to require
the original state
space to be invariant under strong reflection $p^{\mu}\ra -p^{\mu}$.

We cannot have both solutions $|ph\hb_+$ and $|ph\hb_-$ simultaneously since
their bilinear form is
infinite. A simple way  to make sure that only one of the above possibilities
is allowed is to
quantize the particle  in such a way that the spectrum of $v$ is either in
the range $0\leq v<\infty$ or $-\infty<v\leq0$. Such a model is the
$OSp(4,2|2)$-invariant model
considered in \cite{NWe}. This model should therefore  be a canonical model for
the free
spinless relativistic particle (cf. \cite{GoR}).

In the Minkowski treatment leading to \r{413} there is no Lorentz invariant way
to restrict the
original state space to an inner product space  such that the BRST condition
yields a unique
solution, \ie either $|ph\hb_+$ or $|ph\hb_-$. This follows since there is no
Lorentz covariant basis in this case which means that there is no covariant
canonical theory. The
only possibility at our disposal is  an analytic continuation of the Euclidean
treatment above.

\section{The massless  relativistic spin-$\halv$ particle.}
\setcounter{equation}{0}
The standard worldline supersymmetric model for a massless spin-$\halv$
particle \cite{Halv} may be described by the Lagrangian
\be
&&L=\frac{1}{4v}(\dx-i\la\ga)^2-i\halv\ga\cdot\dga
\eel{e:500}
where $\ga^{\mu}$ is an odd Grassmann variable describing the spin degrees of
freedom, and $v, \la$
are Lagrange multipliers. A BRST quantization of \r{500} with dynamical
Lagrange multipliers and
antighosts was considered in \cite{Hwang,MO}. Here we shall quantize this model
on an inner
product space using a graded generalization of the method of ref.\cite{Simple}.
The  BRST charge is
 \be
 &&Q=P^2\eta+P\cdot\ga
c+\pet c^2+\pi\bapet+\kappa\bak \eel{e:501}
where the variables satisfy the following (anti-)commutation relations (the
nonzero part):
\be
&&[\ga^{\mu},\ga^{\nu}]_+=-2\eta^{\mu\nu}, \;\;\; [X^{\mu},
P^{\nu}]_-=i\eta^{\mu\nu},
\;\;\;[\pi, v]_-=-i,\;\;\;[\kappa, \la]_+=1,\nn\\
&&[\pet, \eta]_+=1,\;\;\;[\bapet, \bata]_+=1,\;\;\;[k, c]_-=-i,\;\;\;[\bak,
\bac]_-=-i
\eel{e:502}
where $k,c$ are bosonic ghosts and $\bak, \bac$ the corresponding antighosts,
$\la$ is a fermionic
Lagrange multiplier and $\kappa$ its conjugate momentum. $\eta^{\mu\nu}$ is a
space-like Minkowski
 metric. Notice that
\be
&&[P\cdot\ga, P\cdot\ga]_+=-2P^2
\eel{e:503}
is the algebra of the world-line supersymmetry. In the matrix representation
$\ga^{\mu}$ is turned
into the ordinary Dirac gamma matrices as is shown in the appendix.

By means of a natural generalization of the method of ref.\cite{Simple} it is
easily shown that the
BRST charge \r{501} may be written in the form \r{2}. First one performs a
unitary
transformation to primed variables defined by (the nontrivial part)
 \be
 &&\bapet'=\bapet+\la
c,\;\;\;k'=k+i\la\bata,\;\;\;\kappa'=\kappa-\bata c
\eel{e:504}
where hermiticity and ghost numbers  are preserved. Then one introduces complex
ghosts by
\be
&&\sigma\equiv\halv(\eta'-i\bapet'),\;\;\;\omega\equiv\pet'-i\bata'\nn\\
&&a\equiv\halv(c'-i\bak'),\;\;\;b\equiv ik'-\bac'
\eel{e:505}
satisfying  (the nonzero part)
\be
&&[a, b^{\dag}]_-=1,\;\;\;[\sigma, \omega^{\dag}]_+=1
\eel{e:506}
In terms of these ghost variables $\del$ in \r{2} may be written
\be
&&\del\equiv[a^{\dag}b+\sigma^{\dag}\omega, Q]_-=a^{\dag} D+\sigma^{\dag}\phi
\eel{e:507}
where
\be
&&D=P\cdot\zeta-\la\pi-i\kappa'+a^{\dag}\omega+\omega^{\dag}a,\;\;\;\phi=P^2-i\pi
\eel{e:508}
Notice that  the ghost number operator is
\be
&&N=a^{\dag}b+\sigma^{\dag}\omega-b^{\dag}a-\omega^{\dag}\sigma
\eel{e:509}
Notice also that
\be
&&[D, D]_+=-2\phi,\;\;\;[D, \sigma]_+=a,\;\;\;[D, \sigma^{\dag}]_+=a^{\dag}
\eel{e:510}
imply $\del^2=0$ and  $[\del, \del^{\dag}]_+=0$, and that $D$ and $\phi$
satisfy the original
worldline supersymmetry \r{503}. By means of a bigrading the BRST invariant
states must satisfy
\be
&&\del|ph\hb=\del^{\dag}|ph\hb=0
\eel{e:511}
whose non-trivial solutions we find to be determined by
\be
&&D|ph\hb=\phi|ph\hb=a|ph\hb=\sigma|ph\hb=0
\eel{e:512}
or
\be
&&D^{\dag}|ph\hb=\phi^{\dag}|ph\hb=a^{\dag}|ph\hb=\sigma^{\dag}|ph\hb=0
\eel{e:513}
All other solutions of \r{511} are decoupled zero norm states.
In order to solve these conditions we introduce the transformations ($x$ is any
operator)\cite{Simple}
\be
&&x\;\rightarrow\;e^Axe^{-A}
\eel{e:514}
where
\be
&&A\equiv[\rho, Q]_+=P^2v+iP\cdot\ga\la+k\bak+i\pet\bapet+2i\la\pet
c,\;\;\;\rho\equiv \pet v+k\la
\eel{e:515}
We find then
\be
&&e^A\eta e^{-A}=\eta-2i\la c-i\bapet-\la\bak=2\sigma-2i\la a,\;\;\;e^Ac
e^{-A}=c-i\bak=2a\nn\\
&&e^A(-i\pi)e^{-A}=\phi,\;\;\;e^A(-\la\pi-i\kappa)e^{-A}=D+2\omega a
\eel{e:516}
These expressions imply then that \r{512} is solved by
\be
&&|ph\hb_+=e^{[\rho, Q]}|\Phi\hb
\eel{e:517}
where $|\phi\hb$ satisfies
\be
&&c|\Phi\hb=\eta|\Phi\hb=\pi|\Phi\hb=\kappa|\Phi\hb=0
\eel{e:518}
which are trivially  solved. Since the hermitian conjugation of \r{516} yields
\be
&&e^{-A}\eta e^{A}=2\sigma^{\dag}-2i\la a^{\dag},\;\;\;e^{-A}c
e^{A}=2a^{\dag}\nn\\
&&e^{-A}i\pi
e^{A}=\phi^{\dag},\;\;\;e^{-A}(-\la\pi+i\kappa)e^{A}=D^{\dag}+2\omega^{\dag}
a^{\dag},
\eel{e:519}
eqn \r{513} is solved by
\be
&&|ph\hb_-=e^{-[\rho, Q]}|\Phi\hb
\eel{e:520}
where $|\Phi\hb$ satisfies \r{518}.

The inner products of $|ph\hb_{\pm}$ are
\be
&&_{\pm}\vb
ph|ph\hb_{\pm}=\vb\Phi|e^{\pm2(P^2v+iP\cdot\ga\la+k\bak+i\pet\bapet-2i\la\pet
c)}|\Phi\hb=\nn\\
&&=\vb\Phi|e^{{\pm}2P^2v}e^{{\pm}2iP\cdot\ga\la}e^{{\pm}2k\bak}e^{{\pm}2i\pet\bapet}e^{{\mp}4i\la\pet
c}e^{-4\bak\la\pet}|\Phi\hb=\nn\\
&&=-8\vb\Phi|e^{{\pm}2P^2v}e^{{\pm}2k\bak}P\cdot\ga\la\pet\bapet|\Phi\hb
\mp8i\vb\Phi|e^{{\pm}2P^2v}e^{{\pm}2k\bak}\bak\la\pet|\Phi\hb
\eel{e:521}
The integration over the bosonic ghosts is simple if one uses the rule that
$c,k$ and $\bac,\bak$
must be quantized with opposite choices of state spaces, \ie one with positive
and the other with
indefinite metric states. The integration over $k$ yields then in the first
term $\del(\bak)$ and in
the second $\del(\bak)\bak=0$. Hence, we arrive at ($C$ is a positive constant)
\be
&&_{\pm}\vb
ph|ph\hb_{\pm}=-C\,'\vb\Phi|e^{{\pm}2P^2v}P\cdot\ga\la\pet\bapet|\Phi\hb'
\eel{e:522}
where $|\Phi\hb'$ is equal to $|\Phi\hb$ without the bosonic ghost part. We may
choose
\be
&&|\Phi\hb'=\sum_{\al=1}^4|\psi_{\al}\hb|\al\hb|0\hb_{\kappa}|0\hb_{\eta\bata}
\eel{e:523}
where $|\al\hb$ are spinor states built from the operators $\ga^{\mu}$ (see
appendix). Eqn \r{522}
becomes then
\be
&&_{\pm}\vb
ph|ph\hb_{\pm}=\nn\\
&&=-C\sum_{\al,\beta=1}^4\vb\psi_{\al}|e^{{\pm}2P^2v}P_{\mu}\:\,_{\eta\bata}\vb0|\,
_{\kappa}\vb0|\vb\al|\ga^{\mu}\la|\beta\hb|0\hb_{\kappa}\pet\bapet|0\hb_{\eta\bata}
|\psi_{\beta}\hb
\eel{e:524}
In the appendix it is shown that
\be
&&
_{\kappa}\vb0|\vb\al|\ga^{\mu}\la|\beta\hb|0\hb_{\kappa}=i(\ga^0\ga^{\mu}\ga^5)_{\al\beta}
\eel{e:525}
where $\ga^0\ga^{\mu}$ are Dirac's gamma matrices. Furthermore we may choose
\cite{Mar}
\be
&&_{\eta\bata}\vb0|\pet\bapet|0\hb_{\eta\bata}=i.
\eel{e:526}
 Thus, if we quantize $P^0$ with positive metric states and $v$ with
indefinite ones then \r{524} becomes
\be
&&_{\pm}\vb
ph|ph\hb_{\pm}\propto\int\!d^4p\,\del(p^2)\bar{\psi}(p){p}\!\!\!\slash\ga^5\psi(p)
\eel{e:527}
which is a Lorentz invariant expression. However, it  is not a positive
  norm.  (The left-handed positive energy part and the right-handed
negative energy part have opposite  norms to the right-handed positive energy
part and the
left-handed negative energy part.) It should be emphasized that $\ga^5$ in
\r{527} is forced on us
as soon as odd Grassmann numbers are introduced into the quantum theory. (If
odd Grassmann
numbers are banned from the quantum theory then \r{527} without $\ga^5$ should
be an allowed
possibility. However, this would \eg exclude pseudoclassical path integrals.)
Even within the gauge
fixing procedure of  \cite{Gauge} the norm \r{527} is obtained.

If we instead quantize $P^0$ with indefinite metric states and $v$ with
positive ones, then the
argument leading to \r{416}  applies. Hence, we get
\be
&&_{\pm}\vb
ph|ph\hb_{\pm}\propto\int\!d^4p\,\bar{\psi}(-p^0,\bp)\frac{{p}\!\!\!\slash\ga^5}{p^2}\psi(p^0,\bp)
\eel{e:528}
where $p^0$ is real and Euclidean, \ie $p^2>0$, and ${p}\!\!\!\slash\equiv
ip^0\ga^0+\bp\cdot\bga$.
Also this norm is  not positive definite.

If we instead of $\ga^{\mu}$ in \r{501}-\r{502} had used
fermionic operators $\zeta^{\mu}$  satisfying
\be
&&[\zeta^{\mu}, \zeta^{\nu}]_+=2\eta^{\mu\nu}
\eel{e:529}
then we would not obtain  any factor $\ga^5$ in \r{527} and
\r{528}. The reason for this is obvious since we may make the identification
(see
appendix)
\be &&\zeta^{\mu}=\ga^{\mu}\ga^5
\eel{e:530}
This is a consistent possibility only for massless spin-1/2 particles.
(However, this does  not make
\r{527} positive since the positive and negative energy parts now will have
opposite norms.)

As in the spinless case a canonical theory of the spin-1/2 particle requires a
quantization with an
Euclidean spectrum for the four-momentum and the ranges $0\leq v<\infty$ or
$-\infty <v\leq 0$
for the Lagrange multipliers.

Contrary to the spinless case there is no natural manifestly covariant way to
make the norms \r{527}
and \r{528} positive. The replacement $\ga^{\mu}\ra\zeta^{\mu}$ does not
improve the situation.

\section{The massless  relativistic spin-one particle.}
\setcounter{equation}{0}
The following generalization of the Lagrangian \r{500} was given in \cite{BDH}:
\be
&&L=\frac{1}{4v_1}(\dx-i\sum_{k=1}^2\la_k\ga_k)^2-i\halv\sum_{k=1}^2\ga_k\cdot\dga_k-iv_2\ga_1\cdot\ga_2
\eel{e:601}
Its gauge invariance is an $O(2)$-extended world-line supersymmetry. In
distinction to \r{500} we
have here introduced two odd Grassmannn variables, $\ga_1^{\mu}$ and
$\ga_2^{\mu}$, describing  the
spin degrees of freedom. There are also two odd Lagrange multipliers, $\la_1$
and $\la_2$.
An additional even Lagrange multiplier $v_2$ is introduced as well. That this
Lagrangian describes
a spin-one particle was first proposed in \cite{GT}. It was further treated in
\cite{HPPT,RU}. A
particular interesting feature of this Lagrangian is that there exists no gauge
fixing to the
constraint $\ga_1\cdot\ga_2 $ which means that there exists no  corresponding
regular Lagrangian. A
generalized BRST quantization of \r{601} was performed in \cite{RU} on a
particular state space. Here
we shall perform a standard BRST quantization on an inner product space by
means of the procedure of
\cite{Simple,Gauge}. The BRST charge is \cite{RU}
\be
&&Q=P^2\eta_1+c_1\phi_1+c_2\phi_2+\eta_2B+\pet_1(c_1^2+c_2^2)+2\eta_2(k_1c_2-k_2c_1)+\nn\\&&+\ka_1\bak_1+
\ka_2\bak_2+\pi_1\bapet_1+\pi_2\bapet_2 \eel{e:602} where
\be
&&\phi_i\equiv P\cdot\ga_i,\;\;\;B\equiv i\ga_1\cdot\ga_2
\eel{e:603}
and where $\pet_i,\eta_i$ and $k_i, c_i (i=1,2)$ are fermionic and bosonic
ghosts respectively, and
$\bapet_i,\bata_i$ and $\bak_i, \bac_i (i=1,2)$  the corresponding antighosts.
$\pi_i, v_i$ and
$\ka_i, \la_i$ are the bosonic and fermionic Lagrange multipliers. All
variables satisfy the
following (anti-)commutation relations (the non-zero part)
\be
&&[\ga_i^{\mu},\ga_j^{\nu}]_+=-2\del_{ij}\eta^{\mu\nu}, \;\;\; [X^{\mu},
P^{\nu}]_-=i\eta^{\mu\nu},
\;\;\;\nn\\
&&[\pi_k, v_l]_-=-i\del_{kl},\;\;\;[\kappa_i,\la_j]_+=\del_{ij},
\;\;\;[\pet_i, \eta_j]_+=\del_{ij}\nn\\
&&[\bapet_i, \bata_j]_+=\del_{ij},\;\;\;[k_i,
c_j]_-=-i\del_{ij},\;\;\; [\bak_i, \bac_j]_-=-i\del_{ij}
\eel{e:604}
$\phi_i, B$ and $P^2$ are generators of the $O(2)$-extended world-line
supersymmetry. Their
algebra is
 \be
&&[\phi_i, \phi_j]_+=-2\del_{ij}P,\;\;\;[B, \phi_1]_+=2i\phi_2,\;\;\;[B,
\phi_2]_+=-2i\phi_1
\eel{e:605}

{}From the  treatment of \cite{Simple,Gauge} we expect the general formal
solutions of
$\del|ph\hb=\del\dagg|ph\hb=0$ to be
\be
&&|ph\hb_{\pm}=e^{\pm[\rho, Q]}|\Phi\hb
\eel{e:606}
We shall consider the case when $\rho=\pet_iv_i+k_i\kappa_i$ for which case the
state $|\Phi\hb$
must satisfy \be
&&c_i|\Phi\hb=\eta_i|\Phi\hb=\pi_i|\Phi\hb=\kappa_i|\Phi\hb=0
\eel{e:607}
In order to calculate the norms
\be
&&_{\pm}\vb ph|ph\hb_{\pm}=\vb\Phi|e^{\pm[\rho, Q]}|\Phi\hb
\eel{e:608}
we have to simplify $e^{\pm[\rho, Q]}$. We have explicitly
\be
&&[\rho,
Q]=P^2v_1-i\la_i\phi_i+v_2B+i\pet_i\bapet_i+k_i\bak_i+\nn\\
&&+2i\pet_1(\la_1c_1+\la_2c_2)+2i\eta_2(k_1\la_2-k_2\la_1)+2v_2(k_1c_2-k_2c_1)
\eel{e:609}
Thus,
\be
&&e^{\pm[\rho, Q]}=e^{\pm P^2v_1}e^{\pm\bar{B}}e^{\pm R}
\eel{e:610}
where
\be
&&\bar{B}\equiv-i\la_i\phi_i+v_2B;\;\;\;R\equiv i\pet_i\bapet_i+k_i\bak_i+\nn\\
&&+2i\pet_1(\la_1c_1+\la_2c_2)+2i\eta_2(k_1\la_2-k_2\la_1)+2v_2(k_1c_2-k_2c_1)
\eel{e:611}
The last two exponentials in \r{610} may be further simplified. For
$e^{\pm\bar{B}}$ we find
\be
&&e^{\pm\bar{B}}=\left(1\pm\al(v_2) a+\beta(v_2)
b+\frac{\beta(v_2)}{2v_2}a^2\pm\frac{1}{2v_2}(\al(v_2)-1)ab\right)e^{\pm v_2B}
\eel{e:612}
where
\be
&&a\equiv-i\la_i\phi_i,\;\;\;b\equiv\la_1\phi_2-\la_2\phi_1,\;\;\;a^2=2\la_1\la_2\phi_1\phi_2,\;\;\;ab=2i\la_1\la_2P^2,\nn\\
&&\al(v_2)\equiv\frac{\sinh
2v_2}{2v_2},\;\;\;\beta(v_2)\equiv\frac{1}{2v_2}(\cosh 2v_2-1)
\eel{e:613}
and
\be
&&e^{\pm R}=e^{\pm i\pet_i\bapet_i}e^{\pm k_i\bak_i}G_{\pm}
\eel{e:614}
where
\be
&&G_{\pm}\equiv e^{\pm c+d},\nn\\
&&c\equiv
2i\pet_1(\la_1c_1+\la_2c_2)+2i\eta_2(k_1\la_2-k_2\la_1)+2v_2(k_1c_2-k_2c_1),\nn\\
&&d\equiv
iv_2(k_1\bak_2-k_2\bak_1)-\pet_1(\la_1\bak_1+\la_2\bak_2)+\bapet_2(k_2\la_1-k_1\la_2)
\eel{e:615}
When we now insert these expressions into the right-hand-side of \r{608} we
realize that the terms in
\r{615} involving $\pet_1,\;\bapet_2$ and $\eta_2$ will not contribute to  the
norm. We may therefore
effectively set \be &&c=2v_2(k_1c_2-k_2c_1),\;\;\;d=iv_2(k_1\bak_2-k_2\bak_1)
\eel{e:616}
in \r{615}. We find then
\be
&&G_{\pm}=e^{\halv i(k_1\bak_2-k_2\bak_1)\sinh2v_2}e^{\pm\halv
k_i\bak_i\cosh2v_2}e^{\pm c}
\eel{e:617}
Inserting \r{610} with \r{612}, \r{614}, and \r{617} into \r{608} we find
\be
&&_{\pm}\vb ph|ph\hb_{\pm}=\vb\Phi|e^{\pm P^2v_1}e^{\pm\bar{B}}e^{\pm
i\pet_i\bapet_i}e^{\pm k_i\bak_i}G_{\pm}|\Phi\hb=\nn\\
&&=\vb\Phi|e^{\pm
P^2v_1}e^{\pm\bar{B}}\pet_1\pet_2\bapet_1\bapet_2e^{\halv
i(k_1\bak_2-k_2\bak_1)
\sinh2v_2}e^{\pm k_i\bak_i(1+\halv \cosh2v_2)}|\Phi\hb
 \eel{e:618}

A problem with this $O(2)$-model is that tha gauge generator $B$ cannot be said
to eliminate any
dynamical variable since there exists no gauge fixing to $B$. As a consequence
it is unclear how the
quantization rule in section 2 should be interpreted, \ie whether or not the
spectrum of $v_2$ should
be chosen to be real or imaginary. In the following we shall therefore consider
both cases and derive
under what conditions they lead to finite results. First we notice  that in
order for \r{618} to
have a chance to be finite the eigenvalues of $k_i\bak_i$ must be imaginary
which requires the
bosonic ghosts and antighosts to be quantized with opposite metric states in
accordance with the
second rule in section 2. However, in addition also
$i(k_1\bak_2-k_2\bak_1)\sinh2v_2$ must be
imaginary. This yields the following two possibilities:
 \ben
\item if the spectrum of $v_2$ is real
then $k_1$ and $\bak_2$ must be quantized with the same metric but opposite to
those for $k_2$ and
$\bak_1$. \item if the spectrum of $v_2$ is imaginary then $k_1$ and $k_2$ must
be quantized with the
same metric but opposite to those of $\bak_1$ and $\bak_2$. \een Both these
possibilities are
consistent with the quantization rules of section 2.

In case (1) \r{618} will involve the integral ($v_2$ denotes here the
eigenvalue of the operator
$v_2$)
 \be
&&\int dv_2 d^2k d^2\bak f(v_2,\ldots)e^{\halv
i(k_1\bak_2-k_2\bak_1)\sinh2v_2}e^{\pm
ik_i\bak_i(1+\halv\cosh2v_2)}
 \eel{e:619}
where the function $f(v_2,\ldots)$ does not involve $k_i$ and $\bak_i$. The
integration over $k_i$ and
$\bak_i$ yields here
\be
&&(2\pi)^2\int dv_2 f(v_2,\ldots)\frac{1}{\frac{5}{4}+\cosh2v_2}
\eel{e:620}
In case (2) \r{618} will instead involve the integral
\be
&&\int du_2 d^2k d^2\bak f(iu_2,\ldots)e^{\halv
i(k_1\bak_2-k_2\bak_1)\sin2u_2}e^{\pm
ik_i\bak_i(1+\halv\cos2u_2)}=\nn\\
&&=(2\pi)^2\int du_2 f(iu_2,\ldots)\frac{1}{1+\cos2u_2+\kvart\cos4u_2}
\eel{e:621}
where $iu_2$ is the eigenvalue of the operator $v_2$. Inserting \r{612} into
\r{618} we find
\be
&&_{\pm}\vb ph|ph\hb_{\pm}=(2\pi)^2\,'\vb\Phi|\,_{\pi_1}\vb0|e^{\pm
P^2v_1}|0\hb_{\pi_1}\times\nn\\
&&\times\int
dv_2\left(-i\frac{\beta(v_2)}{v_2}\phi_1\phi_2\pm\frac{1}{v_2}(\al(v_2)-1)P^2\right)\frac{e^{\pm
v_2B}}{\frac{5}{4}+\cosh 2v_2}|\Phi\hb'
\eel{e:622}
in case (1), and
\be
&&_{\pm}\vb ph|ph\hb_{\pm}=(2\pi)^2\,'\vb\Phi|\,_{\pi_1}\vb0|e^{\pm
P^2v_1}|0\hb_{\pi_1}\times\nn\\
&&\times\int
du_2\left(-\frac{\beta(iu_2)}{u_2}\phi_1\phi_2\pm\frac{1}{iu_2}(\al(iu_2)-1)P^2\right)\times\nn\\
&&\times\frac{e^{\pm
iu_2B}}{1+\cos2u_2+\kvart\cos4u_2}|\Phi\hb'
\eel{e:623}
in case (2). We have used the fact that
\be
&&|\Phi\hb=|\Phi\hb'|0\hb_{\kappa}|0\hb_{\pi}|0\hb_{\eta}|0\hb_{\bata}|0\hb_{c}
\eel{e:6231}
from \r{607}, and \r{619}-\r{622}, and the normalization \cite{Mar}
\be
&&_{\kappa,\eta,\bata}\vb0|\la_1\la_2\pet_1\pet_2\bapet_1\bapet_2|0\hb_{\kappa,\eta\bata}=-i
\eel{e:6232}

In order to proceed we must
calculate $\phi_1\phi_2$ and $e^{\pm v_2B}$ and to do that we have to specify
the state $|\Phi\hb'$.
In fact, $|\Phi\hb'$ may be expanded in terms of eigenvalues of $B$ as follows
\be
&&|\Phi\hb'=\sum_{n=-2}^{2}|\Phi,n\hb,\;\;\;B|\Phi,n\hb=2n|\Phi,n\hb
\eel{e:624}
where $n$ is an integer. These eigenstates are easily constructed \cite{RU}:
Define $b^{\mu}$ by
\be
&&b^{\mu}\equiv\halv(\ga_1^{\mu}-i\ga_2^{\mu})
\eel{e:625}
We have then
\be
&&\ga_1^{\mu}=(b^{\mu}+b^{\mu\dag}),\;\;\;\ga_2^{\mu}=i(b^{\mu}-b^{\mu\dag})
\eel{e:626}
and
\be
&&[b^{\mu}, b^{\nu\dag}]_+=-\eta^{\mu\nu}
\eel{e:627}
Hence,
\be
&&B\equiv i\ga_1\cdot\ga_2=-2b\dagg\cdot b-4
\eel{e:628}
The possible matter states are then
\be
&&|\Phi,-2\hb=|A\hb|0\hb,\;\;\;|\Phi,-1\hb=|A_{\mu}\hb
b^{\dag\mu}|0\hb,\;\;\;|\Phi,0\hb=
|A_{\mu\nu}\hb b^{\dag\mu}b^{\dag\nu}|0\hb\nn\\
&&|\Phi,1\hb=|A_{\mu\nu\rho}\hb b^{\dag\mu}b^{\dag\nu}b^{\dag\rho}|0\hb,\;\;\;
|\Phi,2\hb=|A_{\mu\nu\rho\la}\hb
b^{\dag\mu}b^{\dag\nu}b^{\dag\rho}b^{\dag\la}|0\hb
\eel{e:629}
where the vacuum state $|0\hb$ satisfies $b^{\mu}|0\hb=0$. Since $\phi_1\phi_2$
and $B$ commute $\phi_1\phi_2$
does not change the eigenvalues of $|\Phi,n\hb$. This is also easily realized
since the operator $\phi_1\phi_2$
has the following form in terms of $b^{\mu}$
 \be
&&\phi_1\phi_2=i(2P\cdot b\dagg P\cdot b+P^2)
\eel{e:630}

When inserting \r{624} into \r{622} and \r{623} we find that $_{\pm}\vb
ph|ph\hb_{\pm}$ can only be
finite if the range of the spectrum of $v_2$ is finite. If we choose the range
in \r{622} to be
$-L\leq v_2\leq L$  we find
\be
&&_{\pm}\vb ph|ph\hb_{\pm}=(2\pi)^2 \sum_{n=-2}^2\vb
n,\Phi|\,_{\pi_1}\vb0|e^{\pm
P^2v_1}|0\hb_{\pi_1}\times\nn\\ &&\times\left(  a_n(2P\cdot b\dagg P\cdot b +
P^2)+b_nP^2\right)|\Phi, n\hb \eel{e:631}
where
\be
&&a_n\equiv \int_{-L}^L dv_2 \frac{(\cosh 2v_2-1)}{2v_2^2}\frac{\cosh
nv_2}{(\frac{5}{4}+\cosh
2v_2)}=a_{-n}>0\nn\\
&&b_n\equiv
\int_{-L}^L dv_2 \frac{(\sinh 2v_2-2v_2)}{2v_2^2}\frac{\sinh
nv_2}{(\frac{5}{4}+\cosh
2v_2)}=-b_{-n}
\eel{e:632}
In \r{623} the range of $u_2$ must be of the form $-L\leq u_2\leq L$ since
$u_2$ represents an
imaginary eigenvalue of an hermitean operator. We find here \r{631} but with
\be
&&a_n\equiv \int_{-L}^L du_2 \frac{(1-\cos 2u_2)}{2u_2^2}\frac{\cos
nu_2}{(1+\cos2u_2+\kvart\cos4u_2)}=a_{-n}>0\nn\\ &&b_n\equiv
\int_{-L}^L du_2 \frac{(\sin 2u_2-2u_2)}{2u_2^2}\frac{\sin
nu_2}{(1+\cos2u_2+\kvart\cos4u_2)}=-b_{-n}
\eel{e:633}
In the case when $P^{\mu}$ is chosen to have a real spectrum and $v_1$ an
imaginary one we find for
the above two cases:
\be
&&
_{\pm}\vb ph|ph\hb_{\pm}=2(2\pi)^3\int d^4p \del(p^2)\left(
-a_{-1}p_{\mu}A^{*\mu}(p)p_{\nu}A^{\nu}(p)+\right.\nn\\
&&\left.+
4a_0p_{\mu}A^{*\mu\nu}(p)p^{\rho}A_{\rho\nu}(p)-
18a_1p_{\mu}A^{*\mu\nu\rho}(p)p^{\la}A_{\la\nu\rho}(p)\right)
\eel{e:634}
 For the case when $P^0, X^0$ are chosen to have an indefinite
metric basis and imaginary eigenvalues, and $v_1$ a real spectrum with the
ranges $(0,\infty)$ or
$(-\infty, 0)$ we find
\be
&&
_{\pm}\vb ph|ph\hb_{\pm}=(2\pi)^2\int d^4p
\left\{(a_{-2}+b_{-2})A^*(\tp^*)A(\tp)+\right.\nn\\
&&+
2a_{-1}\left(\frac{\tp_{\mu}A^{*\mu}(\tp^*)\tp_{\nu}A^{\nu}(\tp)}{p^2}\right)-(a_{-1}+b_{-1})A^{*\mu}(\tp^*)
A_{\mu}(\tp)+\nn\\ &&
+2a_0\left(-4\frac{\tp_{\mu}A^{*\mu\nu}(\tp^*)\tp^{\rho}A_{\rho\nu}(\tp)}{p^2}+A^{*\mu\nu}(\tp^*)
A^{\mu\nu}(\tp)\right)+\nn\\
&&+36a_1\left(\frac{\tp_{\mu}A^{*\mu\nu\rho}(\tp^*)p^{\la}A_{\la\nu\rho}(\tp)}{p^2}\right)-6(a_1+b_1)A^{*\mu\nu\rho}(\tp^*)
A_{\mu\nu\rho}(\tp)-\nn\\
&&\left.-24(a_2-b_2)A^{*\mu\nu\rho\la}(\tp^*)A_{\mu\nu\rho\la}(\tp)\right\}
 \eel{e:635}
where $\tp\equiv(ip^0,\bp)$ and $p^2\equiv (p^0)^2+\bp^2$. Thus, the measure
$d^4p$ is Euclidean
here.

The above expressions involve too many degrees of freedom. In order to describe
only a spin one
particle we need further projections. An obvious condition to impose is
\be
&&B|\Phi\hb=0
\eel{e:636}
which may be written in a manifestly allowed form \cite{Ax}
$\tilde{B}|\Phi\hb=0$ where $\tilde{B}\equiv[\pet_2, Q]_+$. If  we further
restrict $A^{\mu\nu}$
to be a real field we find from \r{634} (we set $F^{\mu\nu}\equiv A^{\mu\nu}$
where $F^{\mu\nu}$ is
real and give three equivalent forms) \be &&
_{\pm}\vb ph|ph\hb_{\pm}\propto\int
d^4p\del(p^2)p_{\mu}F^{\mu\nu}p^{\rho}F_{\rho\nu}\equiv\nn\\
&&\equiv\halv\int
d^4p\del(p^2)F^{\mu\nu}\left(p_{\mu}p^{\rho}F_{\rho\nu}-p_{\nu}p^{\rho}F_{\rho\mu}\right)\equiv
-\frac{1}{6}\int d^4p\del(p^2)G^{\mu\nu\rho}G_{\mu\nu\rho} \eel{e:637}
where $G_{\mu\nu\rho}$ is a totally antisymmetric field given by
\be
&&G_{\mu\nu\rho}\equiv p_{\mu}F_{\nu\rho}+p_{\nu}F_{\rho\mu}+p_{\rho}F_{\mu\nu}
\eel{e:638}

Eqns \r{413},\r{527} and the second form of \r{637} suggest the general form
\be
&&\vb ph|ph\hb\propto\int d^4pA^*_{\cdot}(p)A^{\cdot}_{ph}(p)
\eel{e:639}
where $A^{\cdot}(p)$  is the off-shell field and $A^{\cdot}_{ph}(p)$ is an on
shell field, \ie a
solution of the equations of motion. We have the following table for massless
particles

\begin{tabular}{|l|l|lr|}\hline
Particle&$A^{\cdot}(p)$&$A^{\cdot}_{ph}(p)$&\\ \hline \hline
spin $0$&$\phi(p)$&$\del(p^2)\phi(p)$&\\ \hline
spin $\halv$&$\psi(p)$&$\del(p^2)p\!\!\!\slash\ga^5\psi(p)$&\\ \hline
spin
$1$&$F_{\mu\nu}(p)$&$\del(p^2)(p_{\mu}p^{\rho}F_{\rho\nu}(p)-p_{\nu}p^{\rho}F_{\rho\mu}(p))$&\\
\hline \end{tabular}\\ \\
For the spin one case we have
\be
&&F^{ph}_{\mu\nu}(p)\equiv\del(p^2)(p_{\mu}A^T_{\nu}(p)-p_{\nu}A^T_{\mu}(p))
\eel{e:640}
where
\be
&&A^T_{\mu}(p)\equiv p^{\rho}F_{\rho\mu}(p)
\eel{e:641}
is a transverse vector field ($p^{\mu}A^T_{\mu}(p)\equiv0$). The first form in
\r{637} may  therefore
be written as
\be
&&\vb ph|ph\hb\propto\int d^4p A^T_{\mu}(p)A^{T\mu}(p)>0
\eel{e:642}

For the Euclidean treatment leading to \r{635} the condition \r{636} and
reality of the field
($A_{\mu\nu}(\tp)\ra F_{\mu\nu}(\tp)$) implies
\be
&&\vb ph|ph\hb=2a_0\int d^4p
\left(-4\frac{\tp_{\mu}F^{\mu\nu}(\tp^*)\tp^{\rho}F_{\rho\nu}(\tp)}{p^2}+F^{\mu\nu}(\tp^*)
F^{\mu\nu}(\tp)\right) \eel{e:643}
When this expression is analytically continued to Minkowski space it agrees
with the path integral
results of \cite{PRi}. (The condition \r{636} is imposed there as well.)

\section{Final remarks.}
\setcounter{equation}{0}
We have performed a detailed analysis of the general formal solutions of a BRST
quantization on
inner product spaces found in \cite{Simple,Gauge}. This analysis has shown that
these formal
solutions can be made exact for all models we have considered and in the
process we have unraveled
the appropriate quantum structure of the state spaces for such a BRST
quantization. We have \eg
found that Lagrange multipliers and the unphysical gauge degrees of freedom as
well as the bosonic
ghosts and antighosts must be quantized with opposite metric state spaces. We
have not concentrated
on the physical implications of the treated models. However, we notice that
\ben
\item it is in general possible to calculate the physical norms in a BRST
quantization of a
relativistic particle model,
\item one may use positivity and the choice of a canonical theory as criteria
to select models.
Hopefully a further analysis of such criteria will make them useful for a
deeper understanding of
second quantization.
\een

After the above analysis was completed we have investigated the implications
for the path integral
quantization \cite{Path}. The results are that the above quantum rules in
general make the solutions
of \cite{Simple,Gauge} consistent with the conventional path integral
expressions \cite{BV}. It is
always the case if the Lagrange multipliers are quantized with indefinite
metric states. Our
Euclidean treatment with positive metric states for the Lagrange multiplier
belonging to the mass
shell condition corresponds to the derivation of propagators when they are
analytically continued to
Minkowski space. In the path integral formulation this derivation requires
instead the introduction
of a convergence factor in the conventional expressions \cite{Teit,PRi}.

\setcounter{section}{1}
\setcounter{equation}{0}
\renewcommand{\thesection}{\Alph{section}}
\newpage
\noindent
{\Large{\bf{Appendix}}}\\
{\bf Properties of the spinor states considered in
section 5.}
 \vspace{5mm}\\

In section 5 in connection with the spin-$\halv$ particle we introduced the
fermionic
hermitian operator $\ga^{\mu}$ which also is a Lorentz vector. It satisfies the
anticommutation
relations \be
&&[\ga^{\mu}, \ga^{\nu}]_+=-2\eta^{\mu\nu}
\eel{e:2.77}
where $\eta^{\mu\nu}$ is a space-like Minkowski metric. These $\ga$-operators
may be written in terms of standard canonical operators, \eg as follows:
\be
&&\ga^0=\theta+\pet_{\theta},\;\;\;\ga^1=\xi+\xi^{\dag}\nn\\
&&\ga^2=i(\xi-\xi^{\dag}),\;\;\;\ga^3=\pet_{\theta}-\theta
\eel{e:2.78}
where $\xi,\xi^{\dag}$ are fermionic oscillators satisfying
\be
&&[\xi,\xi^{\dag}]_+=-1
\eel{e:2.79}
and where $\theta$ and $\pet_{\theta}$ are hermitian fermionic coordinate and
momentum operators satisfying
\be
&&[\theta,\pet]_+=1
\eel{e:2.80}
The compatible state space of $\xi, \theta$ and $\pet_{\theta}$ is spanned by
four
independent states. We may choose
\be
&&|1\hb\equiv-\xi^{\dag}|0\hb|0\hb_{\theta}\nn\\
&&|2\hb\equiv|0\hb\pet|0\hb_{\theta}\nn\\
&&|3\hb\equiv\xi^{\dag}|0\hb\pet|0\hb_{\theta}\nn\\
&&|4\hb\equiv|0\hb|0\hb_{\theta}
\eel{e:2.81}
where $|0\hb$ is the vacuum state for $\xi$ ($\xi|0\hb=0$) and
$|0\hb_{\theta}$ the vacuum state for $\theta$  satisfying
$\theta|0\hb_{\theta}=0$. Their normalization may be  chosen to be
\be
&&\vb 0|0\hb=1,\;\;_{\theta}\vb 0|0\hb_{\theta}=0,\;\;_{\theta}\vb
0|\pet|0\hb_{\theta}=1
\eel{e:2.82}
which implies
\be
&&\vb \al|\beta\hb=g_{\al\beta}
\eel{e:2.83}
where $g$ is the block matrix
\be
&&g\equiv\left(\begin{array}{rl}
0&1\\
1&0
\end{array}\right)
\eel{e:2.84}
Obviously $g^2=\bett$. There is  a metric operator $\eta$ satisfying
\be
&&\vb \al|\eta|\beta\hb=\del_{\al\beta}
\eel{e:2.85}
and   $\eta^{\dag}=\eta$ and  $\eta^2=\bett$.    In fact we have
\be
&&\eta=\ga^0
\eel{e:2.86}
If
we define the matrix representation of $\ga^{\mu}$ by
\be
&&\ga^{\mu}_{\al\beta}\equiv\,'\vb\al|\ga^{\mu}|\beta\hb,\;\;\;'\vb\al|\equiv\vb\al|\ga^0
\eel{e:2.88}
then ($\ga^{\mu}_{\al\beta}$) are Dirac's gamma matrices in the following
representation
\be
&&\ga^0=\left(\begin{array}{rl}
0&1\\
1&0
\end{array}\right),\;\;\;\ga^i=\left(\begin{array}{rl}
0&-\sigma^i\\
\sigma^i&\;\;0
\end{array}\right)
\eel{e:2.90}
where $\sigma^i$ are the Pauli matrices. We may also define a $\ga^5$-operator
by
\be
&&\ga^5\equiv i\ga^0\ga^1\ga^2\ga^3
\eel{e:2.87}
Notice that $(\ga^5)^2=\bett$ and $\ga^{5\dag}=-\ga^5$. Its matrix
representation is
\be
&&\ga^5=i\ga^0\ga^1\ga^2\ga^3=\left(\begin{array}{rl}
1&\;\;0\\
0&-1
\end{array}\right)
\eel{e:2.89}

If one requires the existence of eigenstates to $\ga^{\mu}$ then the
eigenvalues must be odd
Grassmann numbers. As soon as odd Grassmann numbers are introduced one must
prescribe the Grassmann
parity of all states and operators. In \cite{Mar} this analysis was carefully
performed and the
resulting possibilities given.  In the above calculations we have chosen  the
Grassmann parity of the
$\xi$-vacuum in \r{2.81} to be even which is a natural choice. However, notice
that the
$\theta$-vacuum $|0\hb_{\theta}$ is neither even nor odd due to the property
\r{2.82}. In fact it has
a mixed parity, $|0\hb_{\theta}=|0\hb^+_{\theta}+|0\hb^-_{\theta}$. Using the
properties given in
\cite{Mar} we find
 \be
&&_{\kappa}\vb0|\vb\al|\ga^{\mu}\la|\beta\hb|0\hb_{\kappa}=i(\ga^0\ga^{\mu}\ga^5)_{\al\beta}
\eel{e:2.91}
This follows since
\be
&&\la|\beta\hb=\ga^5\overline{|\beta\hb}\la
\eel{e:2.92}
where $|0\hb_{\theta}$ is replaced by
$\overline{|0\hb_{\theta}}\equiv|0\hb^+_{\theta}-|0\hb^-_{\theta}$ in
$\overline{|\beta\hb}$, and
since \cite{Mar}  \be &&_{\theta}\vb 0|\pet_{\theta}\overline{|0\hb_{\theta}}=i
\eel{e:2.93}

If we had started from the hermitian fermionic operators $\zeta^{\mu}$
satisfying
\be
&&[\zeta^{\mu}, \zeta^{\nu}]_+=2\eta^{\mu\nu}
\eel{e:2.94}
instead of \r{2.77}, then we had obtained the matrix representation
\be
&&(\ga^{\mu}\ga^5)_{\al\beta}\equiv\,'\vb\al|\zeta^{\mu}|\beta\hb
\eel{e:2.95}
This follows directly from the above calculations with the identification
$\zeta^{\mu}=\ga^{\mu}\ga^5$. With $\ga^{\mu}$ replaced by $\zeta^{\mu}$ we
would then obtain
\be
&&_{\kappa}\vb0|\vb\al|\zeta^{\mu}\la|\beta\hb|0\hb_{\kappa}=-i(\ga^0\ga^{\mu})_{\al\beta}
\eel{e:2.96}

\end{document}